\documentclass{PoS}
\def\be{\begin{equation}}
\def\ee{\end{equation}}

\title{Universal properties of Wilson loop operators in large N QCD }

\ShortTitle{Universal properties of Wilson loop operators in large N QCD }

\author{\speaker{Rajamani Narayanan}%
\\
        Florida International University, Department of Physics, Miami, FL 33199\\
        E-mail: \email{rajamani.narayanan@fiu.edu}}

\author{Herbert Neuberger\\
        Department of Physics and Astronomy, Rutgers University,
Piscataway, NJ 08855, USA\\
        E-mail: \email{neuberg@physics.rutgers.edu}}

\abstract{Eigenvalues of a Wilson loop operator are gauge
invariant and their distribution
undergoes a transition at infinite $N$ as the size
of the loop is changed.
We study this transition using 
the average characteristic polynomial associated
with the Wilson loop operator.
We derive the scaling function in a certain double scaling
limit for two dimensional QCD and hypothesize that
the transition in three and four dimensional QCD
are in the same universality class. Numerical evidence
for this hypothesis
is provided in three dimensions.
}

\FullConference{The XXVI International Symposium on Lattice Field Theory\\
		 July 14-19 2008\\
		 Williamsburg, Virginia, USA}

\begin{document}
\section{Introduction}

In this talk,
we present the main idea in~\cite{Narayanan:2007dv}
and provide a sampling of the numerical results.

The Wilson loop operator, $W$, is a
unitary operator for SU(N) gauge theories and
can be used as 
a probe of the transition from
strong coupling to weak coupling.
Large (area) Wilson loops are non-perturbative and
correspond to strong coupling.
Small (area) Wilson loops are perturbative and
correspond to weak coupling.

The probe is defined as
\be
{\cal W}_N(z,b,L)=\langle \det (z-W) \rangle
\ee
and is the characteristic polynomial associated with the
operator. 
$W$ is the Wilson loop operator, 
$z$ is a complex number, 
$N$ is the number of colors,
$b=\frac{1}{g^2N}$ is the lattice gauge coupling and
$L$ is the linear size of the square loop.
$\langle\cdots\rangle$ is the average over all gauge fields
with the standard gauge action.

The eigenvalues of $W$ are gauge invariant and so is the
characteristic polynomial. The eigenvalues lie on the
unit circle and all of them will be close to unity for
small loops. The eigenvalues will spread uniformly
over the unit circle for large loops. The characteristic
polynomial exhibits a transition 
at $N=\infty$ when $L\to L_c(b)$. 
This is a physical transition since $L_c(b)$
will scale properly with the coupling, $b$, as one
approaches the continuum limit. 

The scaling function in the double scaling limit can
be derived for two dimensional large $N$ QCD~\cite{Narayanan:2007dv}.
We have numerically shown that three dimensional QCD falls into
the same universality class~\cite{Narayanan:2007dv}.

\section{Two dimensional QCD and a multiplicative matrix model}

Two dimensional gauge theory on an infinite lattice can
be gauge fixed so that the only variables are the individual
plaquettes and these will be independently and identically
distributed.
$W=\prod_{j=1}^n U_j$ where $U_j$s are the transporters
around the individual plaquettes that make up the loop
and $n=L^2$ is equal to the area of the loop.
The measure associated with $U_j$ can be set to
$P(U_j) = {\cal N} e^{-\frac{N}{2} {\rm Tr\ } H_j^2}$
where $U_j = e^{i\epsilon H_j}$ and $\epsilon$ plays
the role of gauge coupling.
The dimensionless area is given by $t=\epsilon^2 n$ 
which
is kept fixed as one takes the
 limit $n\to\infty$ and $\epsilon\to 0$.
This is called the multiplicative matrix model~\cite{janik}.
In the continuum limit,
the parameters $b$ and $L$ get replaced by one parameter,
which is denoted by
$t$ in the model, and the characteristic polynomial
becomes
$${\cal W}_N(z,b,L) \to Q_N(z,t)$$

\section{Average characteristic polynomial}
Using a fermionic representation of the determinant,
one can perform the integration over $U_j$. One
can then perform the integration over the fermionic
variables to obtain the following result for
the characteristic polynomial:
\be
Q_N(z,t)
 =\cases{
\sqrt{\frac{N\tau}{2\pi}}
\int_{-\infty}^\infty d\nu
e^{-\frac{N}{2}\tau\nu^2} 
\left[z-e^{-\tau\nu
-\frac{\tau}{2}
}
\right]^N &  $SU(N)$ \cr
\sqrt{\frac{Nt}{2\pi}}
\int_{-\infty}^\infty d\nu
e^{-\frac{N}{2}t\nu^2} 
\left[z-e^{-t\nu
-\frac{\tau}{2}
}
\right]^N &  $U(N)$ \cr
}\label{qnint}
\ee
Integrating out $\nu$ gives exact continuum polynomial expressions,
\be
Q_N(z,t)
 =\cases{
\sum_{k=0}^N 
\pmatrix{
N\cr k\cr
}
z^{N-k} (-1)^k e^{-\frac{\tau k(N-k)}{2N}} &
 $SU(N)$ \cr
\sum_{k=0}^N 
\pmatrix{
N\cr k\cr}
z^{N-k} (-1)^k e^{-\frac{t k(N+1-k)}{2N}} &
$U(N)$ \cr
}
\ee
\be
\tau=t\left(1+\frac{1}{N}\right)
\ee

\section{Heat-kernel measure}
The result for $Q_N(z,t)$ is consistent with
the heat-kernel measure for $W$:
\be
P(W,\tau ) dW = \sum_R d_R \chi_R (W) e^{-\tau C_2 (R)} dW.
\ee
$R$ denotes the representation,
$d_R$ is the dimension of the representation $R$ and
$C_2(R)$ is the second order Casimir in the representation $R$.
To see this, note that
\be
Q_N(z,t) = \langle \prod_{j=1}^N (z-e^{i\theta_j})\rangle
= \sum_{k=0}^N z^{N-k}
(-1)^k M_k(t).
\ee
If we now take the average, $\langle \cdots \rangle$ over the
heat-kernel measure, we get 
\be
M_k(t) = 
\langle
\sum_{1\le j_1 < j_2 < j_3....< j_k\le 
N}e^{i(\theta_{j_1}+\theta_{j_2}+...+\theta_{j_k})}\rangle
=\langle \chi_k (W) \rangle = 
d_k e^{-\tau C_2(k)} = {N\choose k} e^{-\frac{\tau k(N-k)}{2N}}
\ee 

\section{Zeros of $Q_N(z,t)$}

Since $W$ is a unitary operator, the zeros of
$\det (z-W)$ will lie in the unit circle. One can
show this remains true for $Q_N(z,t)$ when the
gauge group is $SU(N)$. To see this,
we rewrite $Q_N(z,t)$ for $SU(N)$ as
\be
Z_N(z,t)=
Q_N(z,t)(-1)^N e^{\frac{(N-1)\tau}{8}} (-z)^{-\frac{N}{2}}=
\sum_{\sigma_1,\sigma_2,...\sigma_N =\pm\frac{1}{2}} 
e^{\ln(-z)\sum_i \sigma_i}
e^{\frac{\tau}{N} \sum_{i > j} \sigma_i \sigma_j}
\ee
This is the partition function for a spin model with
a ferromagnetic interaction for positive $\tau$.
$ln(-z)$ is a complex external magnetic field.
Therefore, the conditions for Lee-Yang theorem~\cite{Lee:1952ig}
 are fulfilled
and all roots of $Q_N(z,t)$ lie on the unit circle for SU(N).
This is not the case for U(N).

\section{Weak coupling {\sl vs} strong coupling}

The transition from weak coupling to strong coupling can
be intuitively seen using the characteristic polynomial,
$Q_N(z,t)$.
In the weak coupling (small area) limit we have $t=0$ and
$Q_N(z,t)=(z-1)^N$. Therefore, 
all roots are at $z=1$ on the unit circle.
In the strong coupling (large area) limit we have $t=\infty$
and
$Q_N(z,t)=z^N+(-1)^N$.
Therefore, all roots are uniformly distributed on the unit circle.

$Q_N(z,t)$ is analytic in $z$ for all $t$ at finite $N$.
But, this is not the case as $N\to\infty$ and this
leads to a transition from weak to strong coupling in the
$N\to\infty$ limit.

\section{Phase transition in an observable -- 
Durhuus-Olesen transition}

There is a critical area, $t=4$, such that
the distribution of zeros of $Q_\infty(z,t)$ 
on the unit circle has a gap around $z=-1$
for $t < 4$ and has no gap for $t > 4$~\cite{janik,Durhuus:1980nb}.
To see this, we note that
the integral representation (\ref{qnint}) 
is dominated by the saddle point, $\nu=\lambda(t,z)$, given by
\be
\lambda=\lambda(t,z)=\frac{1}{ze^{t(\lambda+\frac{1}{2})}-1}
\ee
With
$z=e^{i\theta}$
and $w=2\lambda+1$, 
$\rho(\theta)=-\frac{1}{4\pi}{\bf Re}\ w$
gives the distribution of the eigenvalues of $W$ on the unit circle.

The saddle point equation at $z=-1$ is
\be
w=\tanh \frac{t}{4}w
\ee
showing that $w$ admits non-zero real solutions for $t>4$.

\section{Double scaling limit}

As $N\to\infty$, one can define a scaling region around
$t=4$ and $z=-1$ by
\be
t=\frac{4}{1+\frac{\alpha}{\sqrt{3N}}};\ \ \ \
z=-e^{\left(\frac{4}{3N}\right)^{\frac{3}{4}}\xi}
\ee
$\alpha$ and $\xi$ are the scaling variables that
blow up the region near $t=4$ and $z=-1$.
We can show that
\be
\lim_{N\rightarrow\infty}
\left(\frac{4N}{3}\right)^{\frac{1}{4}}
(-1)^N e^{\frac{(N-1)\tau}{8}} (-z)^{-\frac{N}{2}}
Q_N(z,t)
=
\int_{-\infty}^{\infty} du e^{-u^4-\alpha u^2+\xi u }
\equiv \zeta(\xi,\alpha)
\ee
which is the scaling function in the double scaling limit
associated with the characteristic polynomial.

We hypothesize that this
behavior in the double scaling limit 
derived for two dimensional large $N$ QCD is universal and
should be seen in the large $N$ limit of
3D QCD, 4D QCD, 2D PCM and other related models.
The modified Airy function, $\zeta(\xi,\alpha)$, is the
universal scaling function.

\section{Large N universality hypothesis}

We can now precisely state the continuum large $N$ universality
hypothesis that can be numerically tested in relevant
models.

Let ${\cal C}$ be a closed 
non-intersecting loop: $x_\mu(s), s\in[0,1]$.
Let ${\cal C}(m)$ be a whole family of loops obtained by
dilation: $x_\mu(s,m)=\frac{1}{m} x_\mu 
(s)$,with $m > 0.$ 
Let $W(m,{\cal C}(*))=
W({\cal C}(m))$ be the family of operators associated
with the family of loops denoted by ${\cal C}(*)$
where $m$ labels one member in the family.
Define
\be
O_N (y,m,{\cal C}(*))=\langle \det (e^{\frac{y}{2}}+e^{-\frac{y}{2}} 
W(m,{\cal C}(*))\rangle
\ee
Then our hypothesis is
\be
\lim_{N\rightarrow\infty} {\cal N}(N,b,{\cal C}(*))
O_N\left(y=
\left(\frac{4}{3N^3}\right)^{\frac{1}{4}}\frac{\xi}{a_1({\cal C}(*))},
m=m_c\left [1+\frac{\alpha}{\sqrt{3N}a_2({\cal C}(*))}\right ]
\right) = 
\zeta(\xi,\alpha)
\ee

\section{Numerical test of the universality hypothesis 
-- 3D large N QCD}

We use standard Wilson gauge action.
The lattice coupling $b=\frac{1}{g^2N}$ has dimensions of
length.
We use square Wilson loops of linear length $L$.
We change $b$ to generate a family of square loops labeled
by $L$.
While doing this, we need to keep 
$0.42 < b < b(V)$ where $V$ is the lattice volume assumed
large enough for large $N$
continuum reduction to hold in the confined
phase~\cite{Narayanan:2003fc,Narayanan:2007ug}.
We use smeared links in the construction of the
Wilson loop operator to avoid corner and perimeter divergences.

\begin{figure}
\vskip 1cm
\centerline{
\includegraphics[width=0.8\textwidth]{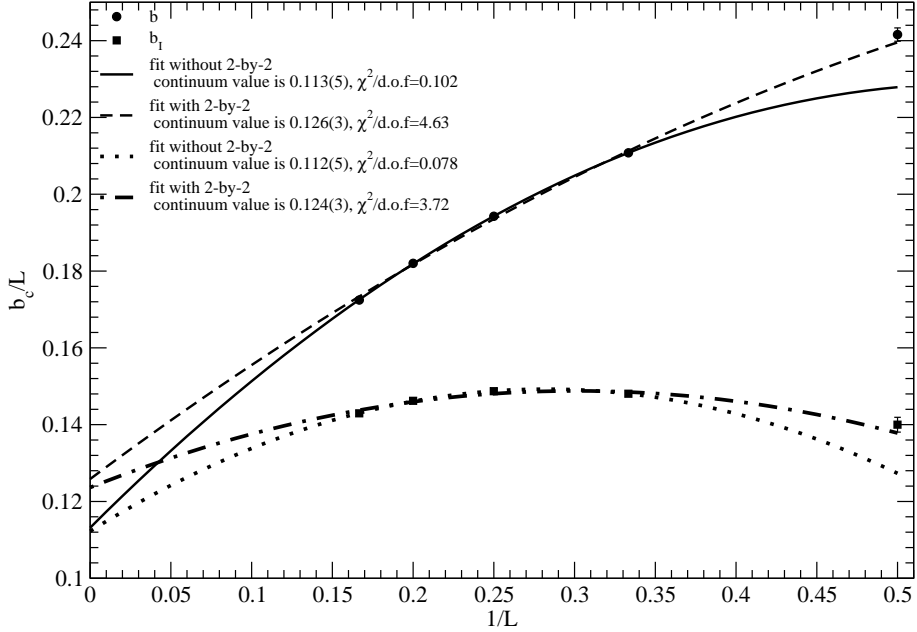}
}
\caption{Extrapolation to the continuum limit of
the critical area.\label{bca}}
\end{figure}

\begin{figure}
\vskip 1cm
\centerline{
\includegraphics[width=0.8\textwidth]{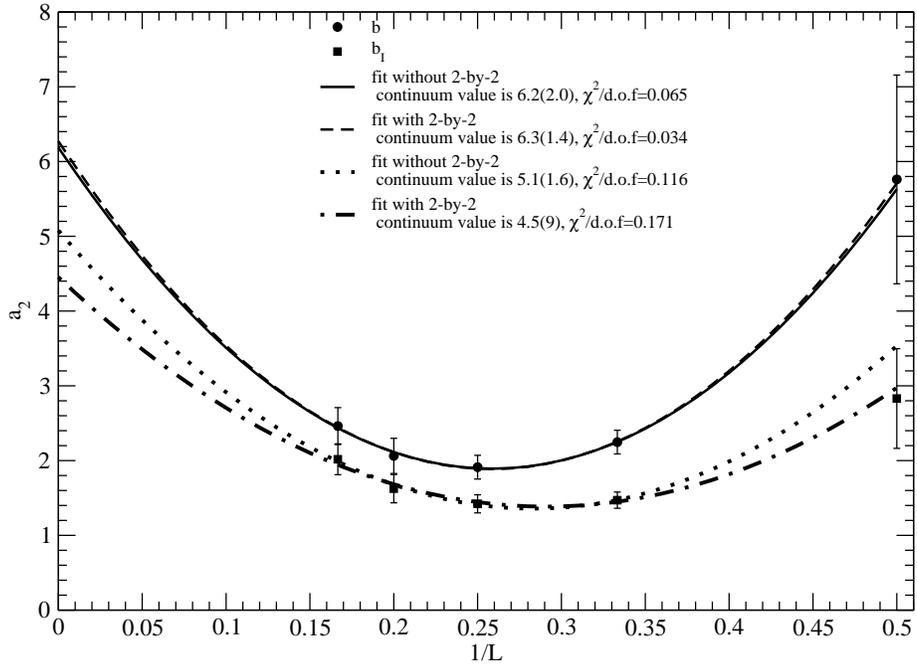}
}
\caption{Extrapolation to the continuum limit of
the parameter matching to the scaled variable, $\alpha$.\label{a2a}}
\end{figure}

\begin{figure}
\vskip 1cm
\centerline{
\includegraphics[width=0.8\textwidth]{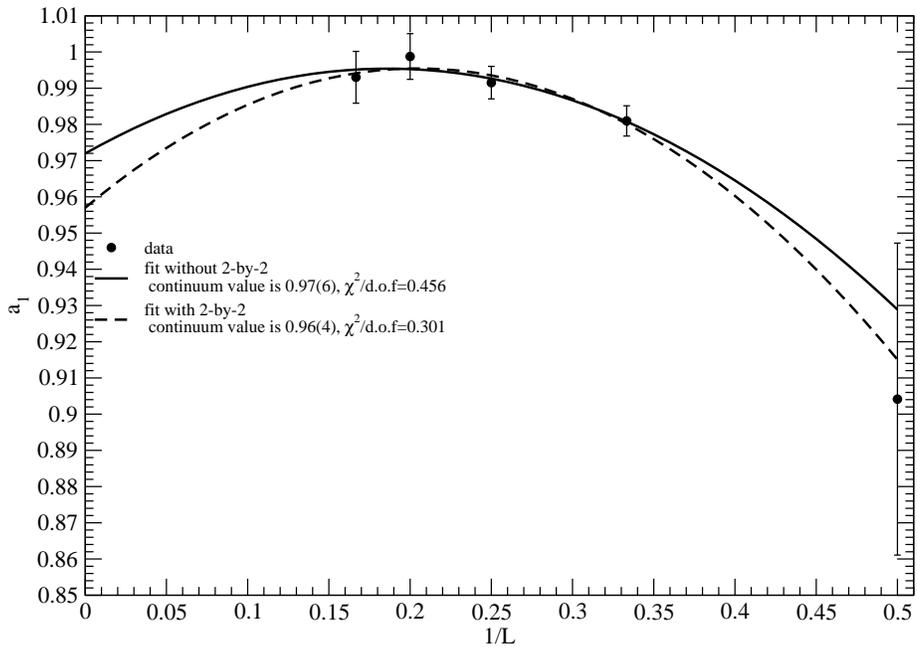}
}
\caption{Extrapolation to the continuum limit of
the parameter matching to the scaled variable, $\xi$.\label{a1a}}
\end{figure}

We obtain $b_c(L)$, $a_1(L)$ and $a_2(L)$ such that
\be
\lim_{N\rightarrow\infty} {\cal N}(b,N)
O_N\left(y=
\left(\frac{4}{3N^3}\right)^{\frac{1}{4}}\frac{\xi}{a_1(L)},
b=b_c(L)\left [ 1 +\frac{\alpha}{\sqrt{3N}a_2(L)}\right ] \right) = 
\zeta(\xi,\alpha)
\ee
This is done
by fixing $N$ and $L$
and obtaining estimates for $b_c(L,N)$, $a_1(L,N)$ and $a_2(L,N)$.
We then take the limit as $N\to \infty$ and
we check that $b_c(L)$, $a_1(L)$ and $a_2(L)$ have
proper continuum limits as $L\to\infty$. The extrapolation
to the continuum limit are shown in Fig.~\ref{bca},
Fig.~\ref{a2a} and Fig.~\ref{a1a}.


\acknowledgments

R.N. acknowledge partial support by the NSF under grant number
PHY-055375 at Florida International University.  
H. N. acknowledges partial support by the DOE, grant \#
DE-FG02-01ER41165, and the SAS of Rutgers University.


\begin{thebibliography}{99}
\bibitem{Narayanan:2007dv}
  R.~Narayanan and H.~Neuberger,
  JHEP {\bf 0712}, 066 (2007)
  [arXiv:0711.4551 [hep-th]].
\bibitem{janik} R. A. Janik, W. Wieczorek, J.\ Phys.\ A:\ Math.\ Gen.
{\bf 37}, 6521 (2004).
\bibitem{Lee:1952ig}
  T.~D.~Lee and C.~N.~Yang,
  Phys.\ Rev.\  {\bf 87}, 410 (1952).
\bibitem{Durhuus:1980nb}
  B.~Durhuus and P.~Olesen,
  Nucl.\ Phys.\  B {\bf 184}, 461 (1981).
\bibitem{Narayanan:2003fc}
  R.~Narayanan and H.~Neuberger,
  Phys.\ Rev.\ Lett.\  {\bf 91}, 081601 (2003)
  [arXiv:hep-lat/0303023].
\bibitem{Narayanan:2007ug}
  R.~Narayanan, H.~Neuberger and F.~Reynoso,
  Phys.\ Lett.\  B {\bf 651}, 246 (2007)
  [arXiv:0704.2591 [hep-lat]].

\end{thebibliography}
\end{document}